# An Approach To Enhance IoT Security in 6G Networks Through Explainable AI

Navneet Kaur, University Of Missouri – St. Louis, USA, nk62v@umsystem.edu
Lav Gupta, University Of Missouri – St. Louis, USA, lgyn6@umsystem.edu

*Abstract*— The evolution of wireless communication has driven many technological advancements, significantly improving connectivity, accessibility, and user experience with each generation. Looking at the 6G framework recently finalized by ITU, the proposed advancements promise unprecedented capabilities, especially for the use cases that are heavily dependent on the Internet of Things (IoT). However, integrating IoT with the 6G infrastructure introduces complex security challenges, many of which remain unexplored. The interconnection of 6G and IoT broadens the attack surface, introducing new vulnerabilities. Also, with the anticipated incorporation of advanced technologies in 6G such as open RAN, terahertz (THz) communication, intelligent reflecting surfaces (IRS), massive MIMO, increased use of AI and disaggregated het-clouds, and many of its proposed use cases like immersive communication, collaborative robotics and native AI support present new security risks while continuing the mutated old ones. Thus, we have new threats relating to AI exploitation, open-source software and increased virtualization along with the existing ones like data manipulation, signal interference, and man-in-the-middle attacks. The complexity and dynamic nature of these technologies can create security blind spots that are difficult to anticipate and mitigate. The 6G standards are expected to be finalized by 2030, with ITU working groups and members focused on aligning security specifications with technological advancements. While many researchers are actively addressing security challenges, significant gaps remain in developing a comprehensive framework to tackle the complex vulnerabilities of integrated IoT and 6G networks. Our research aims to make a meaningful contribution towards addressing some of these gaps by making innovative use of tree-based machine learning algorithm for its ability to manage complex tabular datasets and provide robust feature importance scoring. By employing a cutting-edge dataset that captures 6G network complexities, we implement data balancing to ensure equal representation of attack subcategories. The study also employs interpretability techniques such as SHAP (Shapley Additive Explanations) and LIME (Local Interpretable Model-Agnostic Explanations) for enhanced model transparency, providing both global and local insights. Additionally, our approach involves analyzing the feature importance scores of the model and results of XAI methods, to ensure the alignment between them, cross-validating XAI techniques for consistency, and applying feature elimination to concentrate on the most relevant features, thereby enhancing the model's accuracy. This comprehensive strategy significantly boosts the model's performance and effectiveness in securing IoT within the 6G framework.

## I. INTRODUCTION

Advancements in wireless communication from 3G to 5G have significantly improved technology [1], connectivity, and user experience, but they have also introduced serious security concerns [2]. Despite 5G's intended security enhancements [3], research by GlobalData commissioned by Nokia reveals that nearly three-quarters of 5G network operators have experienced multiple security breaches, leading to network downtime, data leaks, and financial losses. Even after nearly four years since the launch of 5G network, operators report that their defenses remain inadequate against these emerging threats [4]. Considering these concerns, network operators are expected to encounter even greater challenges with the introduction of 6G. Its increased speed, connectivity, and capabilities, coupled with new use cases, may bring about unforeseen vulnerabilities. The integration of massive IoT with 6G could open new avenues for cybercriminals, increasing the number of potential attack vectors and making it more challenging to secure the vast and complex network infrastructure [5].

Moreover, the adoption of various new technologies in 6G and their potential use cases could create blind spots for security experts, leaving room for exploitation [6]. For example, the integration of advanced technologies like Open RAN (Open Radio Access Network) disaggregates traditional network components into modular elements, promoting innovation but also increasing the attack surface and introducing supply chain risks such as compromised hardware or software from multiple vendors, inconsistent security standards, and potential third-party risks [7]. Terahertz (THz) communication, operating at extremely high frequencies, offers benefits such as low latency and high data rates, yet it is susceptible to signal blockage, eavesdropping, and denial-of-service attacks [8]. Intelligent Reflecting Surfaces (IRS), designed to boost signal strength, can be manipulated to redirect signals, leading to communication failures or enabling man-in-the-middle attacks [9]. The use of massive MIMO (Multiple Input Multiple Output) technology, which significantly enhances network capacity, also presents complex security challenges, including pilot contamination and the difficulty of securing a vast array of antennas [10]. The increased use of AI within 6G networks, while optimizing network performance, introduces risks like algorithm manipulation and vulnerability to adversarial attacks [5]. Disaggregated heterogeneous cloud computing (Het-clouds) complicates security further by distributing services across multiple providers, increasing the likelihood of data breaches and insider threats [11].

Anticipated 6G use cases, such as immersive communication (AR/VR), collaborative robotics, and native AI support, also present unique security vulnerabilities. AR/VR technologies can be exploited for data manipulation, where attackers might inject or alter sensory data, leading to misleading or harmful outcomes [12]. Collaborative robotics, relying on real-time communication and coordination, are vulnerable to operational disruptions, hijacking, and other cyber-attacks that could compromise their functionality [13]. Native AI support, deeply integrated into the network, could be manipulated or misused, leading to unintended security consequences and the spread of malicious activities across the network [5].

These use cases, combined with the inherent complexity and dynamism of 6G technologies, highlight the need for robust and adaptive security measures to counter both existing and emerging threats. While AI plays a key role in enabling and securing these new use cases, its lack of transparency can undermine trust, which is essential for the success of advanced IoT-6G applications [14]. Transparency in security systems is absolutely crucial for understanding decision-making processes and ensuring the fairness of security measures [15]. Without clear visibility into AI algorithms, users face significant challenges in validating threat assessments and comprehending the rationale behind security decisions [16]. This lack of transparency severely limits accountability and collaboration, potentially resulting in overlooked threats or misinterpretations [17]. To effectively address these critical issues and protect the evolving 6G and IoT landscape, it is imperative to integrate AI within transparent frameworks and maintain robust human oversight [18].

To ensure transparency in security decision-making for IoT and 6G systems, Explainable AI (XAI) is crucial. XAI helps security professionals understand and manage risks associated with AI-driven decisions [19], clarifies AI's role in security processes, and facilitates accountability for breaches or lapses [20]. This transparency builds trust in AI systems, making security measures more effective and strengthening the overall security framework in the 6G and IoT ecosystem [16].

Despite the significant contributions of existing studies, as outlined in Section 2, several critical challenges remain. Many researchers tend to prioritize Convolutional Neural Network (CNN) models for security applications, often overlooking the superior performance and feature importance that tree-based models can offer [21], especially with complex tabular data [22][23]. Additionally, new datasets that could significantly enhance security analytics in real IoT environments [24] are frequently underutilized. Often, these datasets may not be optimally balanced or may not account for subcategories balance within classes, leading to skewed distributions and unreliable predictions [25]. Explainable AI, which could offer insights into the decision-making processes of complex models [26], has not been thoroughly explored to explain the reasoning behind model predictions. The potential benefits of feature refinement, which could streamline models and enhance accuracy [27], based on insights from Explainable AI (XAI), have not been thoroughly investigated. Moreover, there is a significant lack of comprehensive studies that compare the model predictions with the insights obtained from the XAI methods to verify the consistency and accuracy of the model's prediction. There is also a considerable gap in the research that thoroughly assesses and cross-verifies the results obtained from various XAI techniques to identify inconsistencies or discrepancies among them. Addressing these challenges is crucial for developing robust and effective IoT security solutions, which is the primary motivation behind our research.

This paper tackles the identified challenges by implementing effective strategies in innovative ways to strengthen the security within 6G networks. While the techniques themselves are well established, their customized application to the complex and evolving demands of 6G and IoT environments introduces a novel perspective. By integrating these strategies with the latest advancements in network technology, we worked towards creating a more robust and adaptive security framework. These approaches address specific security gaps and emerging threats in ways that have not been previously investigated. First, we leverage tree-based models (XGBoost, Random Forest, KNN) known for their high performance and ability to handle complex datasets efficiently [21] [22] [23]. This approach provides robust feature importance scoring, allowing us to identify the most relevant features and improve IoT security within the 6G framework. Second, we employ comprehensive datasets that encapsulate the diverse landscape of IoT attacks [24]. This wider scope empowers the model to identify and defend against a broader range of threats, both present and anticipated, thus improving its resilience. Third, we implement robust data balancing technique [28] to ensure that all attack types, including less frequently occurring subcategories, are adequately represented in the training dataset. This strategy enhances the model's ability to learn from a diverse array of scenarios, ultimately improving its generalization capabilities. Fourth, we apply Explainable Artificial Intelligence (XAI) techniques to bolster model transparency and interpretability. This facilitates clear and understandable explanations for the model's predictions, fostering greater transparency and trust in the system. Finally, we incorporate feature elimination technique [29] to refine the model's accuracy by integrating insights from both XAI and the feature importance scores derived from the model. This combined approach demonstrates how these enhancements effectively improve the accuracy of IoT attack detection. Additionally, we verify the model's prediction by comparing the high-impact features identified by the selected model against those highlighted by the XAI methods - LIME and SHAP. This comprehensive approach assesses whether the same features are consistently important across different methods, thereby validating the model's accuracy and trustworthiness in its predictions. Finally, we cross-verified the results of XAI methods against each other to ensure that both models yield consistent outcomes for each data instance used in the analysis. This approach ensures transparency and reliability of the results. Through these advancements, our paper contributes significantly to the development of more robust and trustworthy AI-powered security solutions for the ever-evolving world of IoT. In summary, the contribution of this paper are as follows:

1. Utilizing tree-based machine learning models (Random Forest, XGBoost, and KNN) for our security solution, as they demonstrate superior performance compared to neural networks and then selecting the best performing model for further evaluation and interpretation.

2. Leveraging a new and advanced dataset that accurately capture the complexities of 6G networks to create a robust and practical model, designed to predict and counteract emerging threats, thereby enhancing the efficacy of security solutions in sophisticated network environments.

3. Applying the SMOTE balancing technique to equalize class distributions and address imbalances within subcategories, ensuring reliable predictions across all data categories.

4. Utilizing XAI methods - LIME and SHAP to make the model's predictions more understandable and provide clear explanations for its decisions.

5. Enhanced robustness and accuracy by integrating insights from the model and XAI methods with recursive feature elimination, thereby strengthening the security solution.

6. Performing a comprehensive evaluation to ensure that the high-impact features identified by the model align with those highlighted by XAI techniques, and cross-validate the outcomes of both XAI methods to confirm consistency across benign and attack traffic samples.

7. Conducting a comprehensive feature evaluation to check alignment between the model and XAI results, and cross validating the results of XAI methods, to ensure reliability and transparency.

The rest of the paper is structured as follows. Section 2 reviews related work. Section 3 elaborates on the Proposed Approach. Section 4 details the experiments and performance results of the proposed approach. Finally, Section 5 delves into the findings, conclusions, and directions for future research.

## II. RELATED WORKS

Several studies have effectively used neural networks to detect network traffic vulnerabilities, often preferring CNNs over tree-based models. For instance, in a recent paper [30], the authors evaluate the performance of various deep learning models in detecting cybersecurity attacks within IoT networks. They compare three architectures: Deep Neural Networks (DNN), Long Short-Term Memory (LSTM), and Convolutional Neural Networks (CNN). Another study [31] presents a hybrid oracle-explainer approach for intrusion detection systems (IDS) that uses artificial neural networks (ANNs) and was evaluated on the CICIDS2017 dataset, offering human-understandable interpretations. While these studies make valuable contributions by applying neural networks to intrusion and anomaly detection, there is a growing consensus among researchers that tree based models may provide a more robust and interpretable alternative [21], particularly due to their superior capability in handling complex tabular data [22][23].

Many studies rely on outdated datasets that do not reflect the complexities of modern networks, such as extensive topologies and new attack types [24], limiting their relevance to current network scenarios. For instance, [32] used an XAI framework with SHAP, LIME, CEM, ProtoDash, and BRCG on the NSL-KDD dataset for intrusion detection, while [33] applies SHAP to a multiclass classification problem using the same dataset. Another paper [34] uses XAI with a decision tree algorithm on the KDD dataset to enhance trust management in IDSs. Similarly, the authors in [35] conduct experiments on the NSL-KDD dataset using linear and multilayer perceptron classifiers, providing explanations through intuitive plots. Though these studies offer valuable insights, their findings are constrained because of the use of outdated datasets, reducing their applicability to today's more complex network environments.

While some studies have used new datasets, they often overlook balancing subcategories and do not integrate XAI techniques for explaining the decisions provided by their AI models. For example, the author in [36] uses machine learning algorithms to detect network intrusions in IoT botnet attacks but does not incorporate Explainable AI (XAI) techniques or balance the dataset, leading to potential bias and reduced accuracy in predictions. Similarly, the authors in [37] propose a lightweight deep learning technique for detecting DDoS attacks in IoT environments. Their approach also does not include XAI methods or address dataset balancing. In [38] the author proposes a novel approach to intrusion detection in IoT environments, addressing challenges like resource constraints, security, and privacy, yet it also lacks the integration of XAI and dataset balancing techniques. The study in [39] utilizes tree-based machine learning algorithms for binary, 8-class, and 34-class classification tasks in IoT anomaly detection. The authors balance the dataset and employ a relevant and recent dataset but do not integrate Explainable AI (XAI) techniques to elucidate the reasoning behind the model's predictions, resulting in restricted transparency and interpretability. Another research [40], proposes a hybrid sampling strategy to improve the classification of IoT malicious traffic using tree based algorithms; however, it also does not incorporate XAI methods or balance the dataset. These gaps can lead to the AI models that perform unevenly across different subcategories [41], compromising the overall effectiveness and reliability of the security solutions.

Based on the related work and our own research, it becomes apparent that a successful security framework used in the IoT-6G environment provides better outcomes if it concurrently employs tree based models, uses recent datasets, carries out up to sub-category level balancing of datasets and uses XAI techniques for explaining results and improving the accuracy of predictions.

Furthermore, it is seen that while some researchers utilize explainable AI (XAI) to interpret the results, they do not apply iterative improvements through recursive feature elimination technique, which plays an important role in enhancing model performance model accuracy [27]. The research [42] proposes intrusion detection system (IDS) methods and employed SHAP to interpret the classification decisions of ML models on NetFlow feature sets, including BoT-IoT and ToN-IoT, demonstrating enhanced detection accuracy, but it does not extend these insights to further refine the model through feature elimination technique, which could have led to greater model performance optimization. The paper [43] focuses on explainability in IoT intrusion detection using recent datasets like CICIoT2023 and IoTID20 and applies methods such as LIME and Counterfactual XAI. However, despite employing tree-based models, it does not sufficiently address dataset balancing, compromising the accuracy and fairness of the model's predictions. Additionally, it lacks the inclusion of a feature elimination technique that could enhance model performance.

The authors in [44] propose an IDS using decision trees, random forests, and SVM algorithms. They apply the SMOTE technique to balance the dataset and achieve 96.25% detection accuracy with an ensemble voting classifier. While they use an explainable model for interpreting results, their study relies solely on LIME for explainability and does not incorporate model refinement through feature elimination technique, which could further enhance the model's performance. In [45], a deep neural network is combined with an explainable AI framework (SHAP, LIME, CEM) to enhance network intrusion detection, reaching 82% accuracy on the NSL-KDD dataset. However, the study do not address dataset balancing, and while it employed XAI methods for transparent explanations, it do not compare the insights from different XAI techniques. Additionally, the research utilizes an older dataset and do not incorporate feature elimination techniques. Similarly, [46] utilizes an XGBoost model for network intrusion detection. The authors achieve an accuracy of 93% on the outdated NSL-KDD dataset and uses only the SHAP explanation framework. The study also does not employ any feature refinement technique. Another study [47], uses CICIDS2017 dataset and achieves 90% accuracy. The author only employs the kernelSHAP method to explain the network anomalies and does not use feature elimination techniques to enhance the model's detection capabilities. This study also uses outdated datasets, which may hinder the model's capacity to adapt to current threat landscapes [24].

After an extensive review of the existing literature we can say that to the best of our knowledge, no other publications combines all of these following elements, highlighted in Fig. 1, as we are doing in our work: the application of advanced machine learning techniques, such as tree-based models, which outperform traditional neural networks in handling are more effective with complex tabular data; use of up-to-date datasets that incorporate the intricacies of 5G and 6G networks, ensuring that models are well-suited to the evolving threat landscape; the usage of data balancing technique to ensure all attack types including the rare sub-categories are well represented in the training dataset; the incorporation of transparency through Explainable AI (XAI) methods; the use of feature elimination technique to enhance the detection accuracy and performance of the model; the verification of model's prediction against that of XAI method to assess and evaluate consistency; and cross-validation of the results of different XAI methods to ensure reliable and accurate predictions. This research presents a novel and comprehensive approach that integrates all the complementary techniques to develop accurate, reliable, and adaptable security solutions for IoT deployments in a 6G environment.

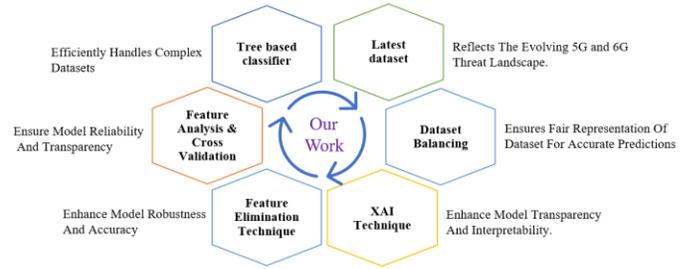

Fig 1. Key Components of Our Work

### III. PROPOSED APPROACH

The proposed model presents an effective application of various methods and techniques for IoT network security. This approach involves multiple stages, as illustrated in Fig 2. The initial stage involves collecting the dataset, where we acquire an IoT attack dataset to train and evaluate the performance of our tree-based models for classifying IoT network traffic. In the second stage, we preprocess the dataset by addressing missing or null values and perform label encoding and data standardization to ensure data consistency. In the third stage, we balance the dataset using the SMOTE technique to address class imbalances and improve the model's ability to generalize across different classes and sub-categories of classes. In the fourth stage, we split the dataset into training and testing sets, using the training set to develop and refine the models and the testing set to evaluate their performance. In the fifth stage, we classify network traffic, using tree-based models (Random Forest, XGBoost, and KNN), to accurately identify and differentiate between benign and malicious network activities. In the sixth stage, we compare the results from these models to identify the best-performing model based on accuracy and feature importance score, for further evaluation and integration with XAI methods. In the seventh stage, we apply XAI techniques— LIME and SHAP— to enhance transparency and to provide insights into the selected model's decision-making process. In the eighth stage, we verify high-impact features identified by the model against those highlighted by XAI methods to assess consistency and evaluate the model's reliability. In the ninth stage, we cross-validate the results obtained from LIME and SHAP to ensure accurate predictions for individual records. Finally, in the tenth stage, we use a recursive feature elimination approach to enhance the model's detection accuracy and refine its predictive capabilities. These techniques help in understanding and optimizing the model's decisions, enhancing performance and reliability, and ensuring robust and trustworthy IoT security for advanced 5G and emerging 6G networks. In the subsequent sub-sections, we provide an in-depth explanation of each of these stages.

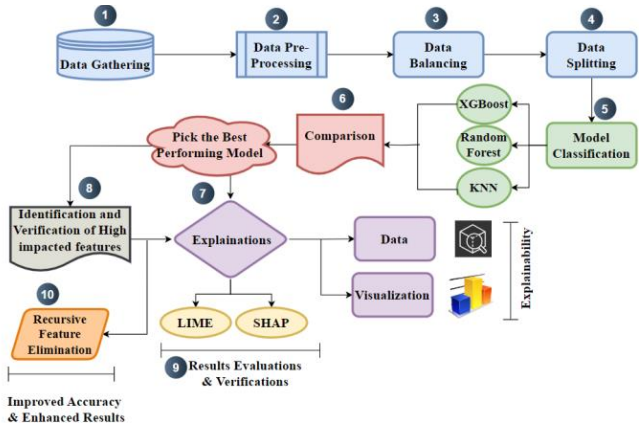

Fig. 2. The Proposed Approach

### 3.1 Dataset Overview and Collection Process

Selecting an appropriate dataset is crucial for providing effective security solutions within the system. As network attacks continuously evolve, relying on outdated datasets may yield less accurate and meaningful results. Therefore, our proposed IoT attack detecting system leverages the latest 'CICIoT2023' dataset from the University of New Brunswick [45]. The dataset comprises 46686579 samples, with 46 attributes. It has one benign class and 33 distinct attack types, categorized into seven primary attack classes - distributed denial of service (DDoS), denial of service (DoS), reconnaissance, web-based, brute-force, spoofing, and the Mirai botnet. The dataset is divided into 169 separate CSV files, each containing a mix of benign and malicious network traffic. Encompassing a wide range of attacks, the dataset provides a robust foundation for developing and evaluating comprehensive security solutions against a diverse threat landscape. Additionally, it is collected from a heterogeneous environment of 105 real IoT devices including smart home devices, cameras, sensors, and microcontrollers, which mirrors the complexity of 5G and 6G networks, making it an ideal dataset for testing and validating advanced security frameworks tailored to these evolving technologies.

### 3.2 Dataset Pre-Processing

Dataset preprocessing involves several crucial steps to enhance data quality and improve model performance. First, the data was converted into Pandas data frames, preparing it for modeling and analysis, which ensures more accurate and efficient workflows. After that, infinite and missing values were addressed to prevent errors and biases in subsequent modeling stages. Duplicate records were removed to ensure the integrity and uniqueness of the data. Categorical labels were converted to numerical representations to enable compatibility with machine learning algorithms. Columns which were containing missing data were eliminated to streamline the dataset and reduce unnecessary complexity. Data normalization was applied to ensure numerical consistency, which enhances model stability and accelerates convergence by bringing all features to a comparable scale. Lastly, features with zero variance were excluded, narrowing the feature set from 46 to 40 attributes, thereby concentrating on the most relevant variables and improving model efficiency. Detailed descriptions of these features are presented in Table I.

This preprocessing ensures that the data is clean and consistent, leading to more accurate and reliable model results.

TABLE I. FEATURE DESCRIPTION

| Feature Num. | Feature | Description |
|---|---|---|
| 0. | flow_duration | Time between the first and the last packet |
| 1. | Header_Length | Header length |
| 2. | Protocol Type | TCP, IP, UDP,ICMP, IGMP, Unknown (Integers) |
| 3. | Duration | Time to live (ttl) |
| 4. | Rate | Pace at which packets are transmitted within a flow |
| 5. | Srate | Rate at which packets are sent out within a flow. |
| 6. | fin_flag_number | Value of fin flag |
| 7. | syn_flag_number | Value of syn flag |
| 8. | rst_flag_number | Value of rst flag |
| 9. | psh_flag_number | Value of psh flag |
| 10. | ack_flag_number | Value of ack flag. |
| 11. | ack_count | Quantity of packets with ack_count in a network flow. |
| 12. | syn_count | Count of packets in a same flow with syn flag set. |
| 13. | fin_count | Count of packets with the FIN flag in a network flow |
| 14. | urg_count | quantity of packets with urg flag in a network flow |
| 15. | rst_count | count of packets where the RST flag is enabled. |
| 16. | HTTP | Identifying if HTTP is the application layer protocol |
| 17. | HTTPS | Identifying if HTTPS is the application layer protocol |
| 18. | DNS | Identifying if DNS is the application layer protocol |
| 19. | SSH | Identifying if SSH is the application layer protocol |
| 20. | TCP | Identifying if TCP is the transport layer protocol |
| 21. | UDP | Identifying if UDP is the transport layer protocol |
| 22. | ARP | Identifying if ARP is the link layer protocol |
| 23. | ICMP | Identifying if ICMP is the network layer protocol |
| 24. | IPv | Identifying if IP is the network layer protocol |

| 25. | LLC | Identifying if LLC is the link layer protocol. |
| --- | --- | --- |
| 26. | Tot sum | Total packets in a flow |
| 27. | Min | Min packet length |
| 28. | Max | Max packet length |
| 29. | AVG | Avg. packet length |
| 30. | Std | Deviation from the mean of packet length within a flow |
| 31. | Tot size | Packet's Length |
| 32. | IAT | Time difference between the consecutive packets. |
| 33. | Number | Count of packets |
| 34. | Magnitude | Square root of the total average packet lengths inside the flow, including incoming and outgoing. |
| 35. | Radius | Square root of the total variation in the inbound and outgoing packet lengths throughout the flow. |
| 36. | Covariance | Connection between the outgoing and incoming packet lengths |
| 37. | Variance | Disparity in packet lengths between incoming and outgoing packets within a given flow. |
| 38. | Weight | Ratio of incoming packets to outgoing packets within a flow. |
| 39. | DHCP | Identifying if DHCP is the application layer protocol |

### 3.3 Data Balancing and Data Splitting

To address class imbalances within the dataset, SMOTE was applied to generate synthetic samples for the minority classes, thereby ensuring a more even distribution of data across all classes. SMOTE is an over-sampling technique used to address class imbalance in datasets by generating synthetic instances of the minority class [40]. Given a minority class instance $x_i$, SMOTE generates synthetic instance $x_{new}$ by linearly interpolating between $x_i$ and its $k$ nearest neighbors.

$$x_{new} = x_i + \lambda \cdot (x_{nm} - x_i) \quad (1)$$

where λ is a random number between 0 and 1 and $x_{nm}$ is one of the $k$ nearest neighbors of $x_i$.

We chose to use the SMOTE technique in our work because it creates new synthetic data points or samples, considering its nearest neighbor, rather than duplicating or removing existing ones, thus preserving all critical information from the majority class. Unlike under-sampling, which can lead to the loss of important data, SMOTE provides a more balanced and informative dataset [25], improving model performance without compromising valuable information. We also converted the problem into a binary classification task, distinguishing between malicious and non-malicious network traffic. The dataset was then split into training and testing sets using an 80-20 ratio for model training and evaluation.

### 3.4 Model Training

For model training, we utilized three tree-based classifiers: XGBoost, Random Forest, and KNN. XGBoost (Extreme Gradient Boosting) is an efficient and high-performance supervised learning algorithm for regression and classification tasks [48]. It builds and combines multiple decision trees sequentially to combine predictive accuracy. Given dataset $D = \{(x_i, y_i)\}_{i=1}^n$ where $x_i$ are input features and $y_i$ are corresponding labels, the prediction $\hat{y_i}$ made by XGBoost for an instance $i$ is given by:

$$\hat{y_i} = \emptyset(x_i) \sum_{k=1}^{k}(f_k(x_i)) \quad (2)$$

Where $K$ is the number of trees, $f_k$ represents the prediction of the $k$-th tree, and $\emptyset(x_i)$ is the final prediction after summing all predictions.

Random Forest is another ensemble learning method that builds multiple decision trees during training and outputs the mode of the classes (classification) or the mean prediction (regression) of the individual trees [49]. Let $T(x, \emptyset_k)$ denote the prediction of the $k$-th tree, for classification, the prediction $\hat{y}$ of the Random Forest is given by:

$$\hat{y} = \frac{1}{K}\sum_{k=1}^{k}(T(x, \emptyset_{k_i}) \quad (3)$$

where $K$ is the number of trees and $\emptyset_k$ represents the parameters of the $k$-th tree.

Another model which we utilized is KNN (K-Nearest Neighbor) is a simple algorithm that classifies data points based on the most common class among their nearest neighbors in the training dataset. It works by finding the k closest data points to a new point and making a prediction based on the majority class or average value of these neighbors [50]. For classification task, the predicted class $\hat{y}$ of an instance of $x$, is determined by:

$$\hat{y} = arg\,max_{y_j} \sum_{i \in N_k(x)} I(y_i = y_j) \quad (4)$$

where $N_k(x)$ is the set of $k$ nearest neighbors of $x$, $y_i$ is the class of $i$-th neighbor and $I$ is the indicator function.

We trained our models using these classifiers to leverage their individual strengths such as XGBoost and Random Forest are known for their robustness in handling complex datasets and capturing intricate patterns [51], while KNN offers simplicity and effectiveness in various scenarios, such as dealing with smaller datasets or when interpretability is crucial [52]. By using these models, we aimed to address the different complexities and challenges inherent in the dataset, leading to a more reliable and accurate classification.

### 3.5 Data Explainability

For our work, we used SHAP and LIME to clearly explain and interpret the model's predictions. SHAP is a method used for explaining individual predictions of machine learning models [53]. It quantifies the contribution of each feature to the prediction by computing Shapley values from cooperative game theory. The SHAP value for feature $i$ in prediction $x$ is expressed as:

$$\emptyset_i(x) = \sum S \subset \{1,2 \ldots p\}\{i\} \frac{|S|!(p-|S|-1)!}{p!} [f(S \cup \{i\}) - f(S)] \quad (5)$$

Here, $\emptyset_i(x)$ denotes SHAP value for feature $i$ at instance $x$, $p$ denotes the total number of features. $f$ is the model's prediction function, $S$ represents the subset of the features excluding $i$, $|S|$ indicates the number of features in subset S and $|S|!(p-|S|-1)!$ signifies the number of ways to select subset $S$

Meanwhile, LIME is another method for explaining individual predictions of machine learning models. It approximates the local behavior of the model around a specific instance by training an interpretable model. LIME approximates the model's prediction function $f$ locally around a given instance $x$ [54]. It does this by minimizing the loss function:

$$\hat{g} = arg\,min_g\, L(f, g,\ \pi_x) + \Omega(g) \quad (6)$$

Here, $\hat{g}$ denotes the interpretable model, $\pi_x$ represents the proximity measure around $x$, $L$ is denoted as loss function and $\Omega(g)$ is a complexity measure of g.

Together, these methods enhance our ability to validate the model's decision-making process and ensure transparency in its predictions.

### 3.6 Recursive Feature Elimination

In our work, we utilized the recursive feature elimination technique (RFE). It is a feature selection method [43] that improves model performance by iteratively removing less relevant or redundant features, based on their impact on predictive power. Let X represent the feature matrix with n features, and $y$ denote the target variable. Start with $X' = X$ (all features included). Develop Machine learning model using $X'$ and $y$. Calculate the feature importance scores, $F = \{f_1, f_2, \ldots \ldots f_n\}$. Identify the least important feature, $f_{min} = argmin\,(F)$. Remove, $f_{min}$ from $X'$. Continue the process until a stopping criterion is reached or the desired number of features are removed, aiming to identify a new feature set that either maintains or improves the score compared to the previous set. The mathematical expression can be written as:

$$Score(X_{new}) = argmin_{x_{new}} Score(X_{new})\, subject\, to\, Score(X_{new}) \geq Score(X_{old}) \quad (7)$$

The purpose of using Recursive Feature Elimination (RFE) is to enhance model performance by identifying and retaining only the most critical features, thereby reducing the noise and improving the model's accuracy.

## IV. EXPERIMENTS AND RESULTS

The experiments are carried out on a GPU-enabled Google Colab environment, utilizing Python 3.7. Several libraries and packages are used for dataset preprocessing, model training, explanations, feature selection, elimination, and visualization. These include Pandas and NumPy for data manipulation, Scikit-learn for machine learning algorithms and preprocessing techniques, TensorFlow and Keras for our learning model development, Matplotlib and Seaborn for visualization, and SHAP and LIME for model explanation and interpretability. This comprehensive toolkit ensures efficient handling of data, robust model training, insightful visualizations and explanations of the results.

### 4.1 Data Handling For Model Training

Due to the dataset's large size and computational constraints, processing all files at once was not feasible. To manage this, we randomly selected 18 CSV data files out of 169 CSV data files, with a total of 100,000 records for training and evaluating our machine learning models, as detailed in Table II. Our main goal is to build a transparent and reliable model using various techniques, including Explainable AI (XAI) methods. By focusing on a subset of the original dataset, we prioritize model interpretability and prediction clarity over comprehensive dataset coverage.

TABLE II. DATASET CATEGORIZATION

| S.num | Class Types | Count | Total Count |
|---|---|---|---|
| 1 | DDoS | 72776 | 97,624 |
| 2 | Dos | 17392 | |
| 3 | Mirai | 5525 | |
| 4 | Spoofing | 1034 | |
| 5 | Recon | 817 | |
| 6 | Web | 52 | |
| 7 | Bruteforce | 28 | |
| 8 | Benign | 2376 | 2376 |

### 4.2 Balancing Using SMOTE

While balancing the dataset, we ensured that the total number of samples in the attack classes equaled that of the Benign class. Additionally, we also ensured that each of the attack class subcategory is also adequately balanced to avoid bias. This approach avoids scenarios where the model might favor majority classes over minority ones, which could negatively impact overall performance and accuracy. As can be seen from Fig. 3(a), the distribution of samples across different categories is uneven, therefore we applied SMOTE to each class and its sub-categories to achieve balanced representation.

For the Benign class, which originally had 2376 samples (refer Table II), we standardized the number to 2100 to ensure better comparability with other classes while maintaining the dataset's robustness and manageability. We then used SMOTE to adjust each attack subcategory, including DDoS (which initially had the highest number of samples), to 300 samples each. This adjustment resulted in a total of 2100 samples for the attack classes, aligning with the count in the benign class. This strategic balancing approach helps maintain a more equitable distribution across classes, improving the model's accuracy in detecting and classifying different types of traffic accurately. The plot illustrates the distribution of each class, providing a visual confirmation of the balanced dataset in Fig. 3(b).

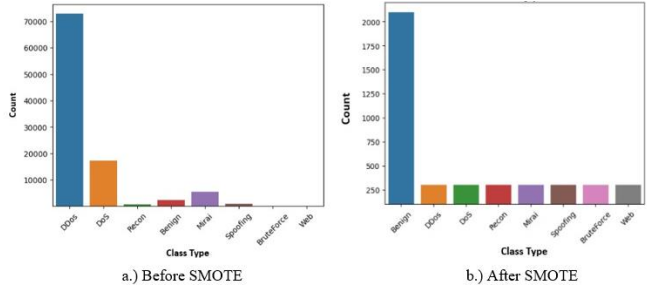

Fig. 3. Visualization Plot of different class types and their counts. Before and After SMOTE

The problem was reformulated into a binary classification task, distinguishing between malicious and non-malicious network traffic. This approach simplifies the classification process by focusing on two primary categories, facilitating more straightforward model training and evaluation, improving overall accuracy and efficiency in detecting security threats. Fig. 4 shows the distribution of the Benign (0) and Attack (1) classes.

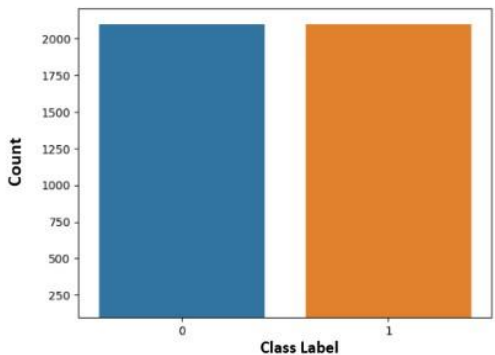

Fig. 4. Distribution of Class Labels – After SMOTE

### 4.3 Model Training Using Tree Based Classifiers

In this section, model training was done, to effectively classify the network traffic as either malicious or benign. We also evaluated the performance metrics of three tree-based models: XGBoost, Random Forest, and KNN, —to compare their results and select the best one. Please note that the results presented are based on an optimal experimentation process designed to maximize insights and practical applicability, rather than solely focusing on accuracy. Our primary goal was to showcase how these techniques can enhance model explainability and transparency, ultimately contributing to a more secure environment. Further refinement and fine-tuning of these processes could be explored to achieve enhanced accuracy.

#### 4.3.1 Implementing XGBoost For Model Training

XGBoost achieved an impressive accuracy of 95.59% on the dataset, showcasing its effectiveness in identifying patterns associated with malicious or benign activity. Fig. 5 displays a feature importance plot that highlights the importance of each feature and feature importance score list that ranks the features from most to least significant. This ranking helps in identifying the most critical features that contribute to the model's predictions and informs decisions on feature retention or exclusion to optimize model performance. It can be noted that Feature 15 (rst_count) is the most influential feature in predicting the outcome, followed by Feature 32 (IAT) and Feature 37 (Variance). Features with a score of zero indicate that they have no impact on output predictions.

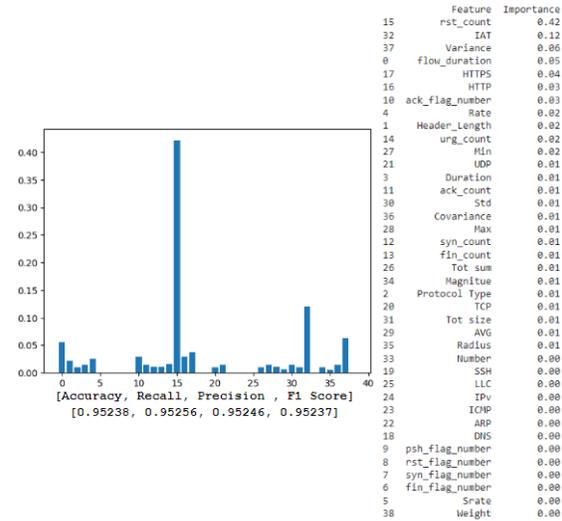

Fig. 5. XGBoost Feature Importance Plot and Feature Importance Score List

#### 4.3.2 Implementing Random Forest for Model Training

Random Forest achieved an accuracy of 94.04%, slightly lower than that of the XGBoost model. However, it is noteworthy that both models identified Feature 15 (rst_count) as the most influential, followed by Feature 32 (IAT) as shown in Fig. 6. This consistency in feature importance highlights the critical role of these features in the model's predictions and underscores their significance in achieving high accuracy.

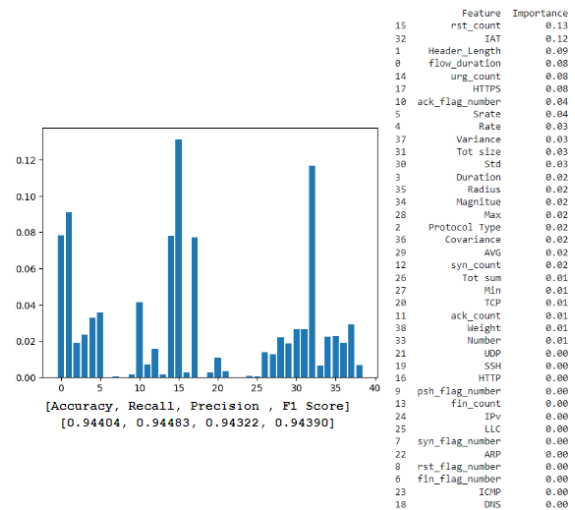

Fig. 6. Random Forest Feature Importance Plot and Feature Importance Score List

### 4.3.3 Implementing KNN For Model Training

KNN (K-Nearest Neighbor) achieved an accuracy of 87.50%, slightly lower than the XGBoost and the Random Forest model. The most influential feature is Feature 32 (IAT), followed by Feature 1 (Header_Length) as shown in Fig. 7. Many features have negative importance scores, suggesting they detract from the model's predictive performance. This indicates that these features might be introducing noise or causing misleading predictions, rather than enhancing accuracy.

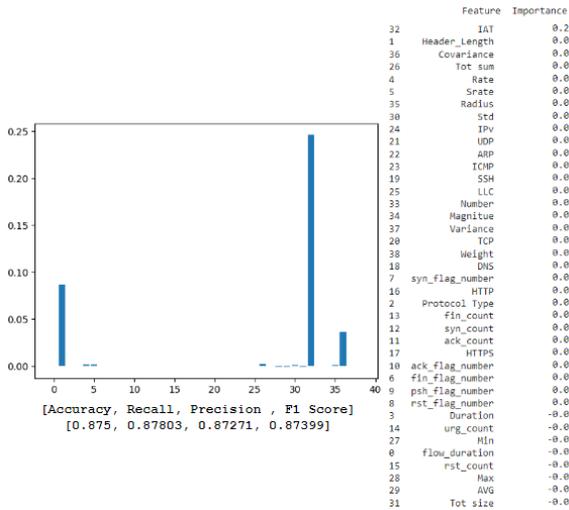

Fig. 7. KNN Feature Importance Plot and Feature Importance Score List.

### 4.4 Comparative Analysis of XGBoost, KNN, and Random Forest Models

This section compares the performance of XGBoost, Random Forest, and KNN to identify the most accurate and reliable model for detecting IoT attacks. Further analysis is carried out on the chosen model to leverage its strengths in achieving optimal performance and mitigating security threats in IoT environment.

### 4.4.1 Comparison Based on Accuracy

Comparative analysis reveals that XGBoost surpasses both KNN and Random Forest in terms of accuracy, as detailed in Table III. Although the XGBoost model achieved an accuracy of 95.59%, which offers a strong baseline, further optimization can be possible. Our objective is to develop a robust model capable of handling the dataset's diversity effectively, avoiding biases towards specific attack subtypes. This baseline model serves as a starting point for in-depth analysis using explainable AI and feature engineering to enhance performance and transparency.

TABLE III. COMPARSION BASED ON ACCURACY

| Models | Accuracy |
|---|---|
| XGBoost | 95.59% |
| Random Forest | 94.04% |
| K-Nearest Neighbor | 87.50% |

### 4.4.2 Comparison Based on Feature Score

A comparative analysis of feature importance scores reveals a convergence between XGBoost and Random Forest, with both models identifying similar influential features. In contrast, KNN highlights a distinct set of important features, as detailed in Table IV.

This consistency between XGBoost and Random Forest makes sense as 'rst_count' (count of TCP reset packets in network traffic) is important for detecting patterns related to connection resets, which can be indicative of scanning or denial-of-service attacks. Similarly, 'IAT (Inter-Arrival Time) is crucial for detecting irregular traffic patterns, such as unusually high or low intervals between packets. For example, a sudden spike in IAT may indicate a Distributed Denial of Service (DDoS) attack, where multiple packets are sent at irregular intervals to overwhelm a network. Conversely, an unusually low IAT might suggest a Brute-Force attack, where rapid, consecutive attempts to breach a system are made. Other features like 'flow_duration' and 'HTTPS' are also essential for detecting network traffic because they provide important context about the nature of the traffic. The feature 'flow_duration' tracks how long a connection persists, helping to identify anomalies such as unusually brief connections, which may signal a denial-of-service (DoS) attack. 'HTTPS' indicates whether the traffic is encrypted; attackers might target unencrypted traffic, while legitimate interactions are typically encrypted. By incorporating these features, the model can more accurately differentiate between standard and suspicious traffic, enhancing its ability to detect both benign activities and potential security threats.

TABLE III. COMPARSION BASED ON FEATURE SCORES

| Models | Highest Influential Feature | Second Highest Influential Feature | Third Highest Influential Feature | Fourth Highest Influential Feature | Fifth Highest Influential Feature | Sixth Highest Influential Feature |
|---|---|---|---|---|---|---|
| XGBoost | rst_count | IAT | variance | flow_duration | HTTPS | HTTP |
| Random Forest | rst_count | IAT | Header_Length | flow_duration | urg_count | HTTPS |
| KNN | IAT | Header_Length | Covariance | Tot sum | Rate | Srate |

Though KNN identified 'IAT' as an important feature, it also predicted 'Header_Length' as the second best. While Header length is a critical component of network packets that can provide valuable insights into the nature of the traffic, including whether it might be malicious, it's not a definitive indicator on its own. It can be a helpful tool in identifying potential threats, when analyzed in conjunction with other packet characteristics.

Taking all these factors into account, we have chosen to focus on the type of model that better capture the most critical aspects of the data for more accurate and reliable predictions. We are using XGBoost due to its high accuracy of 95.59% in classifying network traffic and its ability to consistently identify the same set of influential features as Random Forest, which enhances the reliability of its results. This consistency in feature importance across both the models reinforces our decision to use XGBoost for further analysis and explainability.

4.5 Enhancing Model Accuracy and Transparency with SHAP and LIME

In this section, we provide insights into the model's decision-making process through SHAP and LIME. These explainability techniques help break down the model's predictions, offering a clear understanding of how each feature contributes to the outcome. By making the model's inner workings more transparent, we not only increase trust in its predictions but also enable more informed decisions for system administrators.

4.5.1 SHAP Global Behavior Analysis

We use the SHAP (SHapley Additive exPlanations) method to explain how our selected model arrives at specific classifications for each instance. Fig. 8 highlights the key features identified by SHAP and shows which features contribute the most or least to the model's predictions. Feature names are ordered along the Y-axis in descending order of their impact on the model's predictions, with 'rst_count' being the most influential and 'Protocol Type' the least. The X-axis represents the absolute means of the SHAP values, using distinct colors to represent different classes (0 – Benign Class and 1 – Attack Class). The plot shows that SHAP identifies only 20 out of 40 features as significant and influential, excluding those with minimal or no impact. From the plot we can also infer that 'IAT', 'rst_count', 'urg_count', 'Header Length', and 'flow_duration' are the top five features significantly influencing the model's outcome.

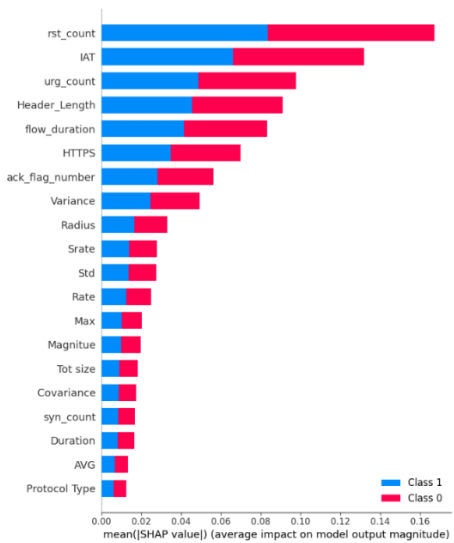

Fig. 8. Summary Plot using SHAP values and Testing Set (Global Explanation)

4.5.2 SHAP Local Behavior Analysis

The SHAP global summary plot offers a broad view of feature importance across the entire dataset, revealing which features are generally most and least influential in model predictions. However, it does not provide details about how these features affect individual predictions such as why the model made a particular decision for a given data point. This is where local analysis becomes essential. Local analysis using SHAP provides detailed, instance-specific insights into feature contributions, thus enhancing model interpretability and reliability.

A. SHAP Summary Plot

For SHAP local explanations, we analyzed two representative instances (Sample 1 and Sample 2) from the testing set. For each instance, we examined how individual features influenced the model's decision— 'Benign' or 'Attack'. We used SHAP local summary plot to visualize and assess the contribution of each feature to the final prediction, providing detailed insights into the model's decision-making process for these specific instances. Fig. 9 illustrates the SHAP plot, where red and blue colors denote high and low feature values, respectively. Features are ranked by their impact on the prediction, with 'rst_count' being the most influential and 'Protocol Type' being the least influential. Plot (A) clearly the 'Benign' class prediction, while plot (B) strongly suggests an 'Attack' classification.

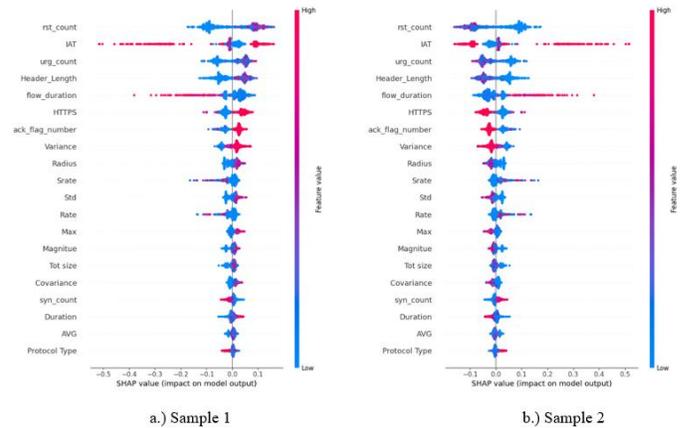

a.) Sample 1     b.) Sample 2

Fig. 9. Local analysis with SHAP a) example of "benign" class prediction b) example of "attack" class prediction

We analyzed the top 5 features—`rst_count`, `IAT`, `urg_count`, `Header_Length`, and `flow_duration`—to explain the model's predictions and revealing their critical roles in distinguishing between benign and malicious attack traffic. We draw the following conclusions from it:

• Low `rst_count` suggests fewer network interruptions, often linked to benign traffic, while high values may indicate anomalies or potential attacks.

• High `IAT` represents longer gaps between packets, typical of benign traffic, whereas low `IAT` could signal frequent, suspicious activity such as DoS attacks.

- Low `urg_count` means fewer urgent packets, which is usual for benign traffic. High values may point to unusual behavior, potentially signaling attacks.
- Low `Header_Length` is common in standard packets, while a high value could indicate the use of custom protocols or malicious activities.
- Long `flow_duration` is characteristic of legitimate connections, whereas short durations may suggest transient, possibly malicious activities.

B. SHAP Force Plot

To delve deeper into individual testing samples (Sample 1 and Sample 2), we employ SHAP force plots. These visualizations offer a granular breakdown of each feature's contribution to a specific prediction. The plot's base value represents the average model output over the training dataset and serves as a starting point for understanding how features influence predictions. Red bars indicate positive contributions, while blue bars represent negative contributions. The length of each bar illustrates the magnitude of a feature's influence, with longer bars reflecting a greater effect. The final prediction displayed at the end of the plot shows the cumulative effect of all feature contributions, starting from the base value.

Fig. 10 shows the SHAP force plot for testing sample 1. The following conclusions can be drawn from it:

- The features 'rst_count', 'HTTPS', 'urg_count', 'Radius', 'ack_flag_number', 'flow_duration', and 'Header_Length' are displayed in red, collectively increasing the prediction score from the base value of 0.49 towards a higher value, supporting the model's classification of the network traffic as 'Benign'.
- Red color signifies a positive contribution to the prediction. For instance, higher values of 'rst_count' might suggest typical session terminations rather than suspicious activities, indicating benign traffic. Also, higher value of 'HTTPS' are usually associated with legitimate and secure communication, reinforcing the notion that the traffic is benign.
- Interestingly, the SHAP force plot for this instance does not highlight 'IAT', previously identified as a crucial feature in the global model. This discrepancy underscores the importance of local explanations, as feature influence can vary significantly across different data points. It suggests that while 'IAT' is generally influential, its impact on this specific case is minimal.

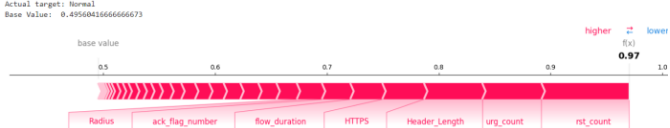

Fig. 10. SHAP Force Plot for testing sample 1 – predicting Class 0 - Benign (Local Explanation)

Fig. 11 shows the SHAP force plot for testing sample 2, the base value is 0.49, and the actual prediction is 'Attack'. The following conclusions can be drawn from it:

- The feature IAT (Inter-Arrival Time) is represented by a long blue bar. This blue color indicates that IAT has a negative contribution to the prediction. Specifically, longer inter-arrival time between packets suggests that the traffic pattern is unusual and potentially indicative of an attack. This negative contribution from IAT pulls the prediction away from benign and towards the 'Attack' classification. In other words, the longer the inter-arrival time, the more it supports the likelihood of the traffic being classified as an attack.
- The other features 'rst_count', 'Header_Length', and 'HTTPS' also have a negative impact, suggesting they support the 'Attack traffic' prediction. This is clearly understood as a high count of reset packets often indicates scanning or denial-of-service attacks, as attackers frequently use these packets to disrupt connections. Also, unusual or varying header lengths can sometimes be indicative of malicious activity, such as attempts to obfuscate the payload or exploit vulnerabilities.
- The features 'Variance', 'ack_flag_number', 'urg_count', and 'flow_duration' are shown in red. This color indicates that these features have a positive contribution to the prediction. A positive contribution means these features support a classification that is less likely to be an attack.

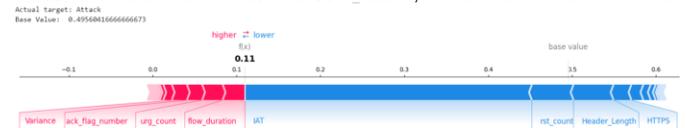

Fig. 11. SHAP Force Plot for testing sample 2 – predicting Class 1 - Attack (Local Explanation)

By comparing these insights with our general understanding of the problem, we can trust that the model is intuitive and is making accurate decisions. For instance, the system is more likely to experience a higher frequency of attacks when there is a substantial number of packets within the same flow and the time difference between packet deliveries is minimal.

4.5.3 LIME Explainer

The LIME results present the prediction probabilities for each class and are divided into three sections:

- The leftmost section displays the prediction probabilities.
- The middle section highlights the most crucial features, with blue representing attributes that support Class 0 and orange representing those that support Class 1. The importance of these features is shown as floating-point numbers.
- The same color scheme is used throughout all sections. The actual values of the top five variables are shown in the final section.

A. LIME Plot - Record Sample 1

Fig. 12 shows the LIME plot for record sample 1, where the prediction is 'Benign traffic. The following conclusions can be drawn:

- The features 'IAT', 'rst_count', 'Std', 'rst_flag_number', and 'flow_duration' are all depicted in blue, indicating that these features negatively impact the prediction, suggesting they push the classification away from 'Benign' traffic.

• Conversely, features like 'Rate', 'HTTPS', 'HTTP', 'fin_count', and 'syn_flag_number' are shown in orange, reflecting their positive contribution to the prediction, which supports the classification of the traffic as 'Benign'. For instance, A higher 'Rate' (rate of packet transmission) is often seen in legitimate network traffic. A high 'HTTPS' and 'HTTP' traffic typically indicate legitimate web traffic, and the presence of 'fin_count' and 'syn_flag_number' within normal ranges suggests regular TCP connection behavior. These positive contributions help reinforce the prediction of the traffic being classified as 'Benign'.

• The 'rst_count' has a very high value of 995.80 which suggests frequent occurrence of reset packets. This high value indicates network issues or potential malicious activity, contributing negatively towards the classification of 'Benign traffic because frequent resets are often associated with attacks or irregular network behavior.

• IAT value is 0.00 which is low suggesting that packets are being transmitted very frequently. Such a value is typically associated with benign traffic as it can indicate regular and uninterrupted communication. in this context, it might indicate a pattern that the model associates with suspicious or anomalous behavior, thus negatively impacting the prediction of Benign traffic. The model might have learned from the training data that such frequent packet transmission correlates more with attack traffic than benign traffic, hence it's shown in blue.

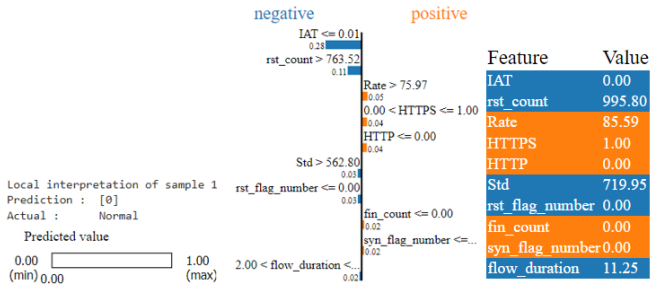

Fig. 12. LIME Plot for Record Sample 1 – predicting Class 0 (Benign)

Notably, the LIME plot identifies only 10 out of 40 features as significant and influential, while SHAP identifies 20 features. This focus selection of LIME is valuable for decision-making as it highlights the most influential factors driving the model's predictions for individual instances. By pinpointing these key features, LIME enables more precise, informed decisions and helps users understand the reasoning behind the model's outputs.

B.  LIME Plot - Record Sample 2

Fig. 13 displays the LIME plot for record sample 2, where the prediction is 'Attack' traffic. The following conclusions can be drawn:

• The features such as 'syn_flag_number', 'rst_flag_number', 'IAT', 'rst_count', and 'ARP' are highlighted in orange, indicating their positive contribution to the prediction of 'Attack' traffic. This means that these features are reinforcing the likelihood of the traffic being classified as malicious. For instance, a higher number of SYN and reset flags might suggest aggressive or disruptive network behavior, while a high inter-arrival time (IAT) and a greater count of TCP resets could point to irregular or suspicious patterns.

• The features such as 'flow_duration', 'Duration', 'Variance', and 'fin_count' are shown in blue, indicating their negative contribution to the prediction of 'Attack'. This suggests that these features are associated with normal traffic patterns, as shorter flow durations and low variance are less characteristic of attack behavior. Additionally, a low fin_count suggests fewer connection terminations or that connections are being closed in a regular and expected manner, which aligns with benign traffic characteristics.

Overall, the plot illustrates how different features influence the model's prediction, with orange features supporting the classification of traffic as an attack and blue features suggesting normal behavior of traffic.

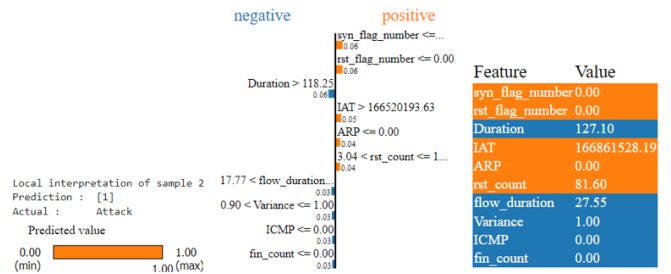

Fig. 13. LIME Plot for Record Sample 2 – predicting Class 1 (Attack)

4.6 Feature Analysis and Cross Validation Using SHAP, LIME, and XGBoost

In this section, we perform a detailed feature analysis by comparing XGBoost's predictions with the explanations provided by SHAP and LIME. We also assess the consistency of SHAP and LIME in identifying key features. This comparison helps validate the crucial factors driving the model's decisions, enhancing transparency and reliability in detecting IoT attacks. Through this analysis, we aim to establish a clear alignment between the model and XAI methods, ensuring robust and explainable predictions.

4.6.2 Feature Analysis by Comparing XGBoost Predictions with XAI Results

When comparing XGBoost results with SHAP and LIME, the analysis involves several key steps to validate the consistency of feature importance across different methods. First, we compare the rankings of feature importance provided by each method, identifying both overlapping and differing features. This comparison helps in evaluating how each technique influences the model's predictions. Next, we focus on features that are consistently highlighted as significant across XGBoost, SHAP, and LIME, which validates their robustness and relevance. Additionally, we cross-verify the important features identified by each method to ensure alignment and consistency in the model's decision-making process. Finally, we analyze how these key features impact the model's predictions and verify the consistency of their effects

across the different techniques. The following analysis can be made:

- Both XGBoost and SHAP highlight 'rst_count' and 'IAT' as highly influential features towards the model's prediction, while 'Protocol Type' is noted as the least influential one. This could be due to the redundancy or overlap of information it provides with other more significant features. For example, flags such as `ack_flag_number`, `syn_count`, `fin_count`, and `urg_count` , which are highlighted as influential features, offer detailed insights into the behavior of TCP connections, often implicitly indicating the protocol in use, rendering 'Protocol Type' feature as somewhat redundant. Moreover, features like `Rate` and `Header_Length` can indirectly capture protocol-related characteristics, while the inclusion of HTTPS/HTTP features further diminishes the standalone importance of 'Protocol Type'. Since these other features collectively capture the key characteristics that 'Protocol Type' would convey, its contribution to the model's predictions becomes less significant.

- The features like 'flow_duration', 'HTTPS', 'ack_flag_number', 'Variance', 'Header_Length', 'Rate', 'Std', 'Max', 'Magnitude', 'Radius', 'Duration', and 'Covariance' are consistently ranked highly by both methods, illustrating their importance. This overlap in feature importance rankings between SHAP and XGBoost reinforces the reliability of these features in the model. Also, by aligning the feature importance results from SHAP and XGBoost, we can confidently assert that both methods agree on the essential characteristics that drive the model's predictions, thus enhancing the transparency and reliability of the model's decision-making process.

- When comparing XGBoost model predictions with LIME results, both methods highlight 'rst_count', 'IAT', and 'flow_duration' as significant features, underscoring their importance in the model's decision-making process. This common recognition underscores the consistent role of these features in predicting traffic types, whether through LIME's localized explanations or XGBoost global model perspective.

In summary, the comparison of XGBoost, SHAP, and LIME consistently identifies key features such as 'rst_count', 'IAT', 'flow_duration', 'Rate', and 'Header_Length' as influential. This consensus among the methods underscores the robustness and reliability of these features in driving the model's predictions, highlighting their crucial role in detecting and analyzing IoT attacks.

4.6.2 Cross Validation of SHAP and LIME Results

In this section, we will present a comparative analysis of the results obtained from LIME and SHAP, focusing on the insights derived from two scenarios: Sample 1 (benign traffic) and Sample 2 (attack traffic). The purpose of this comparative analysis is to evaluate the consistency and reliability of the insights generated by LIME and SHAP across different traffic scenarios. By examining both benign and attack traffic, we aim to ensure that the model's predictions are transparent, trustworthy, and applicable to a range of real-world situations, thereby improving decision-making and model interpretability. The analysis below highlights the comparison of LIME and SHAP for both benign and attack traffic, demonstrating their respective insights and consistency across different traffic types.

- For benign traffic, both LIME and SHAP identify several key features that contribute to the classification. LIME highlights `Rate`, `HTTPS`, `HTTP`, `fin_count`, and `syn_flag_number` as positive indicators of benign traffic, reflecting regular web and TCP behavior. In contrast, SHAP emphasizes `rst_count`, `HTTPS`, `urg_count`, `Radius`, `ack_flag_number`, `flow_duration`, and `Header_Length` as significant for benign traffic, showing a broader alignment with typical benign patterns.

- For attack traffic, LIME points to `syn_flag_number`, `rst_flag_number`, `IAT`, `ARP`, and `rst_count` as positive indicators, suggesting the presence of attack patterns. It also identifies `Duration`, `flow_duration`, `Variance`, `ICMP`, and `fin_count` as negative indicators, implying deviations from attack characteristics. SHAP similarly uses `IAT`, `rst_count`, `Header_Length`, and `HTTPS` as negative features supporting the attack prediction, while `Variance`, `ack_flag_number`, `urg_count`, and `flow_duration` are the indicating features that are less likely associated with attacks.

In summary, both methods emphasize key features such as 'rst_count' and 'IAT' for differentiating between benign and attack traffic and identify 'rst_count', 'HTTPS', and 'flow_duration' as influential for benign traffic. For attack traffic, both provide complementary view where, SHAP highlights features like 'IAT', 'rst_count', 'Header_Length', and 'HTTPS', while LIME points out 'syn_flag_number', 'rst_flag_number', 'IAT', and 'ARP' as significant. Additionally, it is notable that for a given sample, the predictions made by SHAP and LIME explanations consistently provide the same Class Type. This alignment between SHAP and LIME in predicting the same class type for a given sample underscores the reliability and coherence of the model's behavior, reinforcing the trust in the model's interpretability and decision-making process.

4.7 Recursive Feature Elimination

This section outlines a systematic approach to feature selection and evaluation by integrating XGBoost with XAI techniques. This innovative combination not only identifies crucial features but also enhances model accuracy, showcasing an effective method for improving predictive performance in complex datasets. Initially, we analyze XGBoost insights and eliminate features with zero scores, indicating minimal impact on predictions. We then use XAI techniques to identify both significant and insignificant features. By iteratively refining the XGBoost model with these selected features, as depicted in Fig. 14, we achieved a significant accuracy improvement from 95.59% to 97.02%. The feature importance score list confirms that Feature 6 (rst_count) is the most influential in predicting the output, followed by Feature 12 (IAT), which were previously identified as high-ranking features.

This demonstrates that concentrating on the most relevant features can substantially boost model performance, resulting in more accurate and dependable predictions.

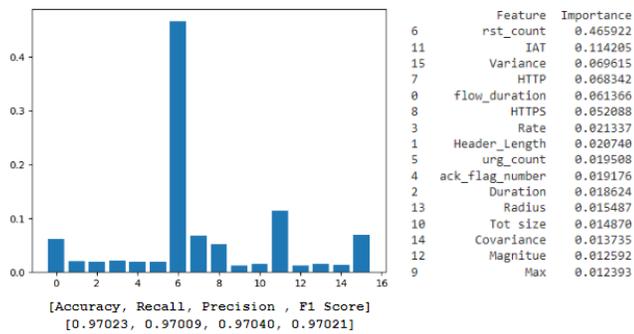

Fig. 14. XGBoost Feature Importance Plot and Feature Importance Score List– After gathering insights from XAI.

## V. CONCLUSION

Ensuring transparency and trust in AI systems is vital for foster a safer and more secure future, particularly in the fast-evolving domain of IoT security. As transformative technologies like 6G reshape our digital landscape and the proliferation of IoT devices accelerates, gaining insight into AI's decision-making processes becomes crucial for addressing emerging challenges. Without clear visibility into how AI systems operate, we risk exposing critical infrastructures to vulnerabilities and threats that could undermine global security. The ability to decode and understand AI's inner workings is not just an advantage—it's a necessity for maintaining the integrity, resilience, and reliability of the interconnected world we are building.

In our study, we assess three tree-based models (Random Forest, XGBoost, and KNN) using the latest CICIoT 2023 dataset which consists of live traffic from a wide range of IoT devices. To ensure reliable predictions, we use the SMOTE technique, to balance all subcategories of attacks within the main classes. We then evaluate these models based on accuracy and feature importance scores to assess their performance and identify which model delivers the most reliable predictions. By examining both the reliability of feature importance scores and the output accuracy, we select the best-performing model to ensure a robust and suitable choice for the next stage of our evaluations. We use XAI techniques, specifically SHAP and LIME, to gain deeper insights into the model's decision-making process. By analyzing both attack and benign data samples, we verify that the predictions from the XGBoost model align with those from SHAP and LIME. This assessment ensures consistency and thereby validates the model's effectiveness. Additionally, we also cross-verify the explanations provided by SHAP and LIME to ensure consistency for the same data sample. Leveraging these insights, we apply recursive feature elimination to boost detection accuracy by removing less significant features, particularly with zero importance scores in the model, and focus on the ones identified as impactful by XAI techniques. This strategic refinement results in a significant accuracy improvement, reaching 97% compared to the 95% accuracy achieved by XGBoost without the use of explainability techniques. Overall, the integration of XAI techniques into our modeling process enhances understanding of the model's inner workings and boosts its accuracy and reliability. These advancements enable administrators to implement more resilient security measures tailored to specific threats and vulnerabilities, thereby enhancing overall system security and protecting against cyber threats and attacks in the growing network.

We believe this research is pivotal as it promotes deeper exploration into integrating XAI techniques with AI models to enhance decision-making processes, reveal hidden patterns, and address emerging challenges in evolving 5G and 6G environments. Furthermore, it will enhance our capabilities in threat detection and network security, better equipping us to manage the complexities of integrating extensive IoT systems with cutting-edge wireless technologies.